# Etching of Graphene Devices with a Helium Ion Beam

Max C. Lemme[a], David C. Bell[b,c], James R. Williams[a,b], Lewis A. Stern[d], Britton W.H. Baugher[e], Pablo Jarillo-Herrero[e], Charles M. Marcus[a,*]

[a]Department of Physics, Harvard University, Cambridge, MA 02138, USA

[b]School of Engineering and Applied Sciences, Harvard University, Cambridge MA, 02138, USA

[c]Center for Nanoscale Systems, Harvard University, Cambridge MA, 02138, USA

[d]Carl Zeiss SMT Inc., Peabody MA 01960, USA

[e]Department of Physics, Massachusetts Institute of Technology, Cambridge, MA 02139, USA

* Address correspondence to marcus@harvard.edu

**Abstract**

We report on the etching of graphene devices with a helium ion beam, including *in situ* electrical measurement during lithography. The etching process can be used to nanostructure and electrically isolate different regions in a graphene device, as demonstrated by etching a channel in a suspended graphene device with etched gaps down to about 10 nm. Graphene devices on silicon dioxide ($SiO_2$) substrates etch with lower He ion doses and are found to have a residual conductivity after etching, which we attribute to contamination by hydrocarbons.

Graphene, a stable two-dimensional carbon crystal, has attracted great interest recently as a model system for fundamental physics as well as for possible nanoelectronics applications.[1-3] Many experiments in the field are targeted at graphene devices where artificial confinement in one or two dimensions produces nanoribbons or quantum dots. Typically, such structures are on the ~5 to 50 nanometer scale and have been fabricated by electron beam lithography followed by reactive ion etching,[4-7] by chemical means such as thermally activated nanoparticles[8] or unfolding of carbon nanotubes.[9-11] While these methods are suitable to produce devices near the atomic limit, they also have significant shortcomings. Reactive ion etching typically erodes the resist mask creating disordered graphene edges. Chemical methods can result in irregular shaped and distributed flakes poorly suited for integrated device applications. It has further been proposed to etch graphene at the nanoscale with a focused electron beam.[12] This method, however, requires suspending graphene on specific transmission electron microscope grids, making it difficult to perform simultaneous electrical measurements.

Helium ion microscopy (HeIM) has recently been introduced as high-resolution imaging technology for nanoscale structures and materials.[13-15] In this work we use a helium ion microscope (Zeiss ORION) as a lithography tool to controllably modify electrical properties of graphene devices. We further demonstrate *in situ* electrical measurement during lithography. The HeIM is particularly well suited for this purpose because it produces a high-brightness, low-energy-spread, sub-nanometer size beam. The microscope benefits from the short de Broglie wavelength of helium, ~ 100 times smaller than the corresponding electron wavelength. This gives the beam an ultimate resolution of 0.5 nm or better,[14] making it an attractive tool for precision lithography of graphene devices. While process details are published elsewhere,[16] this letter focuses on the modification of electrical properties of graphene. Fig. 1 shows a schematic of a graphene field effect transistor as used in this work. Note that for some experiments, the $SiO_2$ substrate was removed prior to measurements to obtain a suspended graphene device (see Methods section). The inset in Fig. 1 shows a photograph of a chip carrier inside the HeIM as used for *in-situ* measurements.

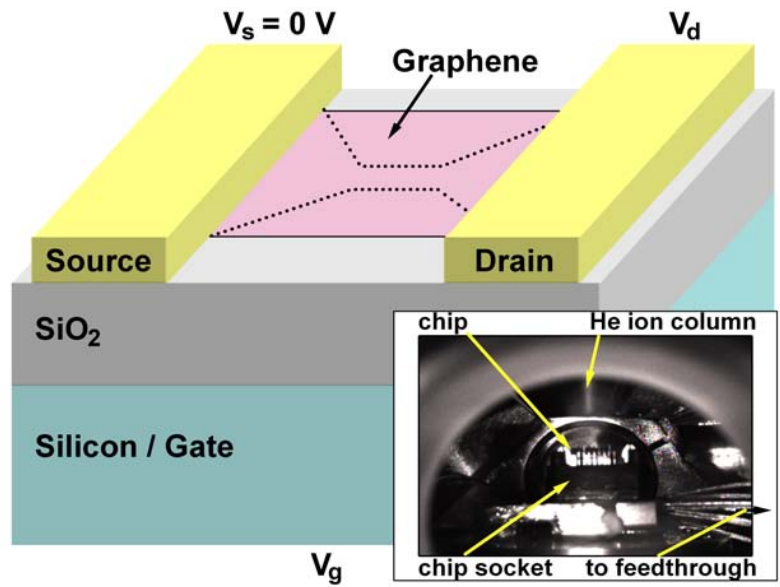

**Fig. 1:** Schematic of a graphene device. Inset: Photograph of the microscope chamber with installed chip.

## Results and Discussion

A suspended graphene device with a length of ~ 150 nm and a width of ~ 1.5 μm, shown in the HeIM microscope image in Fig. 2a, was He ion etched by sequential imaging in high resolution. The graphene was exposed to the He ion beam at a field of view of 2 μm x 2 μm and an image size of 2048 x 2048 pixels, which resulted in a pixel spacing of ~ 1 nm. The dwell time was chosen to be 50 μs resulting in an effective line dose of 0.8 nC/cm. Fig. 2b shows such a high resolution image, expanded and labeled to distinguish the suspended graphene from the underlying $SiO_2$ and the chromium (Cr) / gold (Au) contacts. Fig. 2c shows a sequence of images taken under these conditions (number 1-13, where image 1 is identical to Fig. 2b). The red circle indicates the region of the graphene flake where etching occurred initially. Each scan with the He ion beam resulted in an increase of etched area. After thirteen scans, the dwell time, and hence the image quality, was increased to 500 μs, equivalent to a line dose of 8 nC/cm, still not sufficient to completely etch the device (Fig. 2c, 14). These images indicate that removal of edge atoms is favorable over atoms within in the graphene crystal.

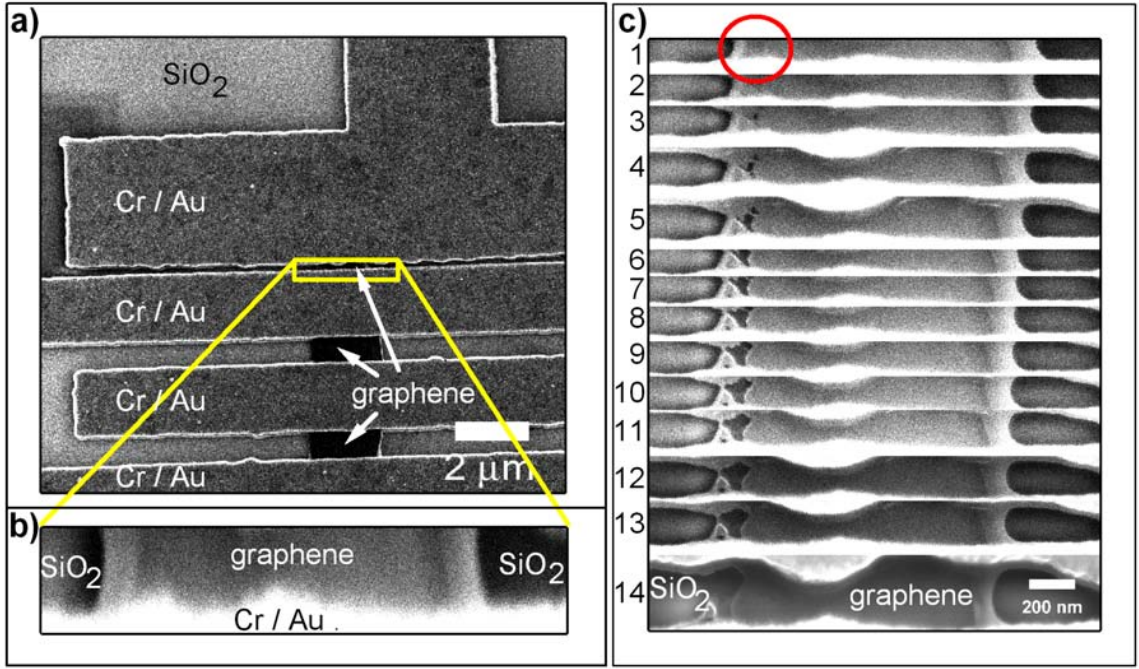

**Fig. 2:** a) HeIM image of suspended graphene devices. The yellow box indicates the area that was subsequently imaged and etched in high resolution. b) High resolution image used to etch graphene. c) Sequence of images of progressive etching of a suspended graphene sheet. Image 1 corresponds to Fig. 2b. The red circle indicates the area where etching occurred initially (color online)

The remaining graphene film was etched using live scanning mode with a 100 nm to 10 nm field of view. Here, etching was confirmed via the live screen image. A resultant cut with minimum feature sizes in the 10 nm range is shown in the HeIM image in Fig. 3a. The gap was measured with DesignCAD software after importing the original image.

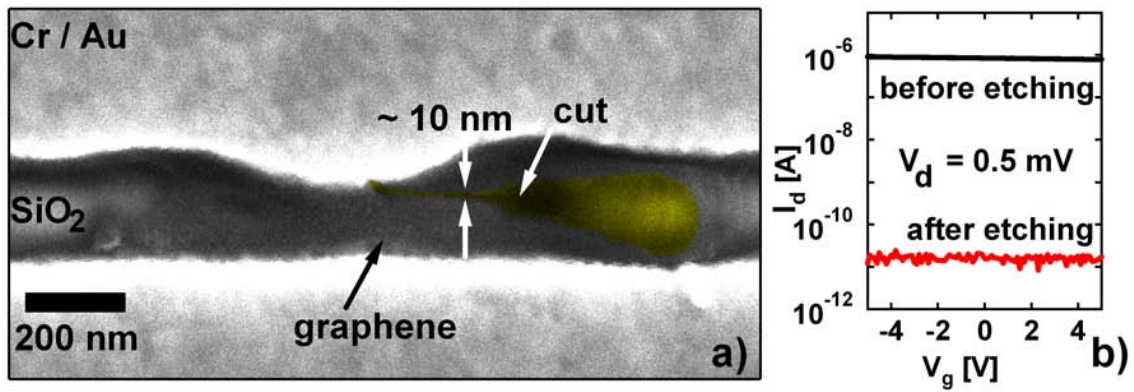

**Fig. 3:** a) HeIM image (with false color) of a suspended graphene device after etching with minimum feature sizes of about 10 nm (color online). b) Electrical measurement of the device before and after etching.

After etching a trench across the entire graphene flake, the device was removed from the HeIM and its drain current was measured as a function of back gate voltage (Fig. 3b, $V_d = 0.5$ mV, note that the gate voltage range is limited in suspended graphene devices,[18,19] and hence $I_d$ changes little with $V_g$). The current dropped to about 15 pA, compared to 1 µA prior to etching. While the latter is typical for a functional graphene device of the given dimensions, the post-etching value corresponds to the noise level of the measurement setup. Adjacent, non-imaged devices made from the same graphene flake showed conductivity similar to the investigated device prior to imaging. These results confirm that the graphene was etched successfully using the He ion beam.

Next, the drain current of a graphene device on $SiO_2$ substrate was measured inside the He ion microscope while part of it was exposed to the ion beam. A field of view of 1 µm x 1 µm was chosen, indicated by the yellow box in Fig. 4a. After about 150 seconds the current saturated, indicating complete etching of the graphene inside the field of view (Fig. 4b). At this point the imaging window was moved to the next part of the device in the direction of the white arrow in Fig 4a. The current was again monitored until it saturated. A beam current of 1 pA, dwell time of 3 µs, and pixel spacing of ~1 nm allowed us to estimate a suitable He ion line dose for etching graphene on SiO$_2$: 1.5 nC/cm.

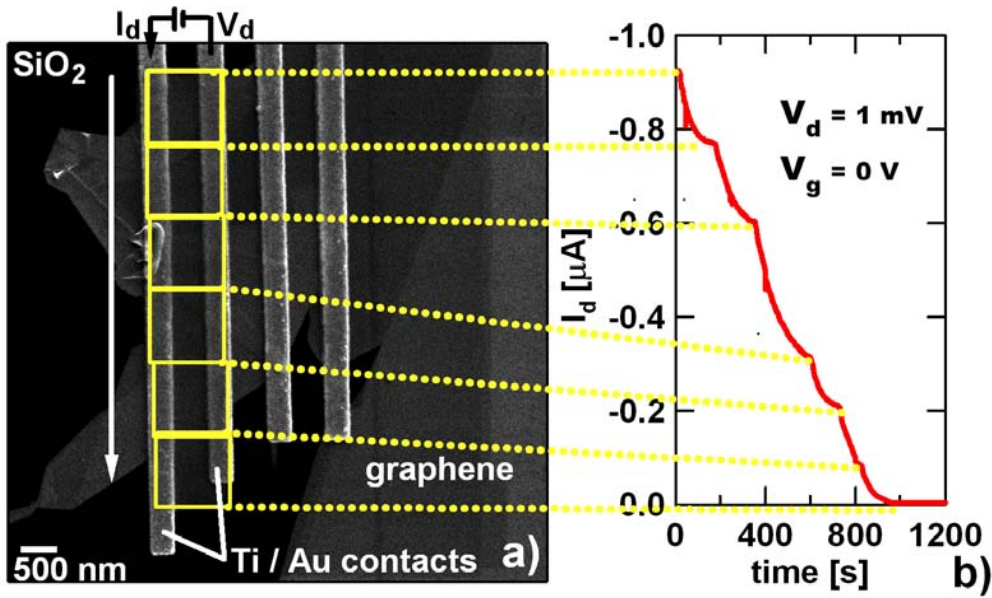

**Fig. 4:** a) HeIM image of a graphene device. The boxes indicate the field of view used for etching. The window was subsequently moved in the direction of the arrow. b) Drain current vs. time of exposure of the graphene device. The etching window was moved as the current saturated.

A residual drain current of about 4 nA was measured after etching the entire device, which could not be reduced further by subsequent He ion beam exposure. We attribute this residual conductivity to contamination of the SiO$_2$ surface with hydrocarbons.

**Conclusions**

We have demonstrated etching of graphene devices with a helium ion beam. Suspended graphene has been etched conclusively, with minimum feature sizes in the 10 nm range. Graphene on SiO$_2$ was etched with a lower dose compared to suspended graphene. However, these devices showed a residual conductivity attributed to contaminants on the surface. Helium ion etching can be considered an alternative nanofabrication method for suspended graphene devices and, if contamination issues can be solved, graphene on $SiO_2$ substrates.

**Methods**

Graphene was deposited onto ~300 nm of silicon dioxide on degenerately doped silicon by mechanical exfoliation[17], similar to the method described by Novoselov et al.[1] Next, mono- and few layer graphene flakes were identified with an optical microscope. Contacts to the graphene were defined by conventional electron beam lithography, followed by evaporation of chromium/gold (3 nm/150 nm) and titanium/gold (5 nm/40 nm). Suspension of the graphene sheet was obtained by wet etching of the underlying $SiO_2$ in diluted HF, followed by critical point drying. All devices were measured in a standard field effect transistor (FET) -like configuration, with the evaporated contacts acting as source and drain, and the doped substrate as a gate electrode (Fig. 1). The drain current $I_d$ through the flake is then measured as a function of gate voltage $V_g$ for a constant drain voltage $V_d$. Electrical data of suspended devices were made outside the microscope, before and after He ion etching, using two Keithley 2400 source meters in a Desert Cryogenics probe station at a pressure of ~$5 \times 10^{-3}$ mbar. The second set of graphene devices on $SiO_2$ substrate were wirebonded to chip carriers and placed in a chip socket inside the Helium ion microscope to enable *in-situ* electrical measurements (inset in Fig. 1). These were taken at a pressure of ~$1 \times 10^{-6}$ mbar with an Agilent 4155B parameter analyzer connected to the device via a vacuum feedthrough. All measurements were taken at room temperature.

**Acknowledgement**

M. Lemme acknowledges the support of the Alexander von Humboldt foundation through a Feodor Lynen Research Fellowship. Research also supported in part by the NRI INDEX program. The authors thank S. Nakaharai for fruitful discussions regarding the process.